\documentclass{article}
\usepackage{natbib}
\usepackage{amsmath}
\usepackage{graphicx}
\usepackage[margin=1in]{geometry}
\usepackage{mathrsfs}
\usepackage{listings}

\begin{document}
	\title{Competing risks joint models using R-INLA}
	  \author{Janet van Niekerk,
		Haakon Bakka and
		H{\aa}vard Rue \\
		CEMSE Division, 
		King Abdullah University of Science and Technology,
		Saudi Arabia}
	\maketitle
\abstract{The methodological advancements made in the field of joint models are numerous. None the less, the case of competing risks joint models have largely been neglected, especially from a practitioner's point of view. In the relevant works on competing risks joint models, the assumptions of Gaussian linear longitudinal series and proportional cause-specific hazard functions, amongst others, have remained unchallenged. In this paper, we provide a framework based on R-INLA to apply competing risks joint models in a unifying way such that non-Gaussian longitudinal data, spatial structures, time dependent splines and various latent association structures, to mention a few, are all embraced in our approach. Our motivation stems from the SANAD trial which exhibits non-linear longitudinal trajectories and competing risks for failure of treatment. We also present a discrete competing risks joint model for longitudinal count data as well as a spatial competing risks joint model, as specific examples.
}




	\section{Introduction}
	
	Time to event data is observed in its simplest form, as the time when a particular event happens or when the monitoring process halts (censoring time). A competing risks model \citep{hoel1972,moeschberger1971,altshuler1970} arises in the case where patients are at risk for one of multiple events, i.e. multiple mutually exclusive outcomes. The occurence of an event precludes all other events, so that each observational unit is at risk for each of the $C$ events, until one event occurs. After this, the risk of the other events are all zero. There are mainly two modeling approaches in this respect, the subdistribution \citep{fine1999} and cause-specific hazards approaches \citep{prentice1978}. In the subdistribution approach the data enters the model through a predictor term in each of the cumulative incidence functions, the cumulative probability of experiencing the specific event. On the other hand, the cause-specific approach incorporates the data through a predictor term for each of the cause-specific hazard functions. Debates surrounding the choice between the two approaches are still ongoing, with the consensus being that the cause-specific hazards approach should be chosen for causal inference of covariate effects.\\ \\
	In most situations, additional data is available in the form of repeated measurements over time constituting a longitudinal data set. The joint modeling of the competing risks data and the longitudinal data \citep{hu2009,elashoff2007} is often necessary since the missing longitudinal observations are not missing at random but rather due to the occurrence of one of the competing events. Ignoring this phenomena, will bias the longitudinal estimates. If the treatment efficacy is being evaluated, then the joint modeling enables us to evaluate the treatment effect in both endpoints simultaneously. Unobserved heterogeneity in the individuals due to some latent effect over time can be successfully accommodated through the joint modeling.\\ \\
	Computational frameworks for competing risks, have been developed in recent years such as the \textit{CFC} package \citep{cfcpkg} and the \textit{cmprsk} package \citep{cmprskpkg}. The existence of these frameworks is essential for catalyzing the use of competing risks models in practice.\\
	For joint modeling of competing risks and longitudinal data the R package \textit{joineR} \citep{joinerpkg} can be used. This package implements the model developed by \citet{williamson2008}. This model is a cause-specific Cox proportional hazards regression model (for two competing events) and a linear mixed effects model with an assumed latent Gaussian association process. An accelerated failure time (AFT) model, a non-Gaussian distribution for the longitudinal model component or spatial data thus excludes the use of \textit{joineR}.\\
	Practitioners depend largely on the availability of wieldy computational frameworks to ease the computational burden of implementing such joint models. On the flip side, too many different developments could confuse the potential users and exacerbate the hesitation to applying competing risks models. This is especially true when more complex models are needed and new packages are developed for each additional complexity. This is cumbersome for users since many different packages should be familiarized for common use. The exigency for a more cohesive approach to competing risks joint models is clear. We believe that we address this deficiency in this paper.\\ \\
	To achieve our aim, of presenting a cohesive computational framework for competing risks joint models, we are spurred to characterize competing risks models within the class of latent Gaussian models. This characterization enables us to use the integrated nested Laplace approximation (INLA) method proposed by \citet{rue2009}. Our rationale for this endeavor is the computational efficiency and ability to incorporate various modeling elements available in the accompanying \textit{R-INLA} package \citep{inlapkg}. The INLA framework has been used in various fields of application since its inception. In survival analysis, \citet{martino2011} provided the details for basic survival models and \citet{alvaro2013} presented an approach for competing risks models using R-INLA, while \citet{van2019} presented the case of joint survival longitudinal modeling. Competing risks joint models with frailties \citep{martino2011}, spatial structures \citep{bakka2018} and non-linear covariate effects \citep{lindgren2008} with non-Gaussian longitudinal submodels are all natural possibilities within our framework. \\ \\
	In the SANAD trial \citep{marson2007}, the aim was to evaluate the efficacy of various treatments for seizure control in patients with epilepsy. Two treatment failure types, unacceptable adverse effects (UAE) and inadequate seizure control (ISC) were considered as competing events. From exploratory data analysis the non-linear mean trajectory of the titration dosage over time for the two different treatments is clear and also noted by \citet{williamson2008}. A piecewise linear model with two different slopes (with the changepoint at time $500$) was used by \citet{williamson2008} as an attempt to properly model this phenomena. We, however, propose a fully nonparametric spline component to capture this non-linear behavior over time in the form of a random walk order two model. This type of non-linear competing risks model can be formulated as a latent Gaussian model which can then be fitted using the established R-INLA framework for accurate and efficient computation. This endeavor resulted in better results for the SANAD trial as well as the development of a class of latent Gaussian competing risks joint models for various other realistic data situations such as longitudinal count data, spatial data and correlated competing risks to name a few.\\ \\
	In Section \ref{joint} we present an introduction and necessary details of competing risks joint models in general which is then formulated as latent Gaussian models in Section \ref{seccmp}. We also provide two specific examples of competing risks joint models which might arise in practice to give the reader an intuition for the proposed methodology. We apply the proposed method to the SANAD trial in Section \ref{SANAD} and a simulation study for a discrete competing risks joint model (longitudinal count data) is presented in Section \ref{simstudy}. Finally, we conclude with a discussion in Section \ref{disc} on the proposed method and some exciting possibilities thereof.
	
	\section{Joint models for competing risks and longitudinal data}\label{joint}
	The applications of joint models have increased over the last few years. Essentially, a joint model is a statistical method to model two or more dependent endpoints/responses. Joint models in spatial statistics are known as coregionalization models \citep{schmidt2003,krainski2018} where there is dependency between multiple spatial fields. In survival analysis, the use of joint models to infer joint survival longitudinal models is very popular. In these cases, joint modeling is fundamentally correct since the time to event process and the longitudinal data per individual is inherently dependent, since it is governed by the same biological process. Although joint models are  popular in survival analysis with one event of interest, the potential for their application in more complicated survival models has yet to be explored. In this paper, we focus on competing risks joint models with various realistic features like mutiple events of interest, spatial random effects and non-linear longitudinal trajectores. To this end, we present cause-specific competing risks joint models, with proportional or non-proportional cause-specific hazard functions and non-Gaussian or Gaussian longitudinal series as latent Gaussian models. This venture allows the use of the fast and accurate INLA methodology \cite{rue2009} for Bayesian inference of the model.
	\subsection{Cause-specific competing risks joint model}
	The competing risks joint model is used to do inference for the biological process that underpins the longitudinal and competing risks data. For simplicity we assume $C$ competing risks and one longitudinal marker in this section, even though our approach can be used for various longitudinal markers.\\
	For the longitudinal marker $\pmb{y}$, we formulate a linear predictor $\eta^L$ and vector of hyperparameters, $\pmb{\theta}^L$ for $N$ individuals with a total of $N^L=N_1+N_2+...+N_N$ observations. Note that we do not make any distributional assumptions, hence denote the likelihood of $\pmb{y}$ as
	\begin{equation}
	\pi_L(\pmb{y}|\pmb{\eta}^L,\pmb{\theta}^L)=\prod_{l=1}^{N_L}\pi_L(y_l|\eta^L_l,\pmb{\theta}^L).\label{longlik1}
	\end{equation}
	\\ \\
	As before, denote the cause-specific hazard function for cause $j, j=1,...,C$ at time $t$ as 
	\begin{equation*}
		h_{j}(t|\eta^j,\pmb{\theta}^j),\label{cshfunction}
	\end{equation*} where $\eta^j$ is the linear predictor. The vector of hyperparameters $\pmb{\theta}^j$ is associated with the cause-specific hazard function, for example the baseline hazard parameters for the Cox proportional hazard model or the shape parameter of the Weibull accelerated failure time (AFT) model.  The likelihood function for the observed event times $\pmb{t}=\{t_1,...,t_N\}$ with event indicators $\pmb{d}=\{d_{ij}\},i=1,...,N,j=1,...,C$ is
	\begin{equation}
	\pi_S(\pmb{t}|\pmb{\eta},\pmb{\theta})=\prod_{i=1}^N \prod_{j=1}^C h_{ij}(t_i|\eta^j_i,\pmb{\theta}^j)^{d_{ij}}\exp\left(-\int_0^{t_i}h_{ij}(u|\eta^j_i,\pmb{\theta}^j)du\right),\label{survlik1}
	\end{equation}
	where $d_{ij}$ is an indicator if individual $i$ died from cause $j$. 
	The linear predictors, $\eta^j$ and $\eta^L$, have an additive structure of fixed and random effects as follows:
	$$\eta=\pmb{\beta}\pmb{Z}+\sum_{k=1}^{K}f_k(u_{ik})$$ where $f_k(u_{ik})$ is the $k^{th}$ random effect of covariate $u_{ik}$ for individual $i$. This formulation allows us to include spatial fields, spline components and frailties within our approach to joint competing risks models. The form of the linear predictor facilitates the contextualization of this joint competing risks model as a latent Gaussian model.\\
	\section{Competing risks joint models as Latent Gaussian models}\label{seccmp}
	In this section we briefly present the details of latent Gaussian models (LGMs) and the integrated nested Laplace approximations (INLA) method for approximate Bayesian inference. If we can formulate competing risks joint models as LGMs and consequently use INLA for the inference, then we can fit these types of models very efficiently and with complex modeling components.
	\subsection{Latent Gaussian models and the integrated nested Laplace approximations method}
	Latent Gaussian models is a specific subset of hierarchical Bayesian additive models. 
	This class comprises of well-known models such as mixed
	models, temporal and spatial models amongst many others. An LGM is defined as a model
	having a specific hierarchical structure, as follows:\\ \\
	\begin{enumerate}
		\item {The observed responses, $\{y_i\}$,
			are conditionally independent based on the likelihood hyperparameters
			$\pmb{\theta}_1$ and the linear predictors $\{\eta_i\}$. (Note that non-Gaussian distributions as well as multiple distributions for subsets of the responses, can be assumed.)}
		\item {The linear predictor
			is formulated as follows:
			\begin{equation}
			\eta_i=\beta_0+\pmb{\beta}\pmb{X}+\sum_{k=1}^Kf_k(u_{ik})+\epsilon_i
			\label{additive predictor}
			\end{equation}
			where $\pmb{\beta}$ represent the linear fixed effects of the
			covariates $\pmb{X}$, $\pmb{\epsilon}$ is the unstructured random effects
			and the $\{f_k(.)\}$'s represent unknown
			(non-linear) functions of the covariates $\pmb{u}$. The
			unknown functions $\{f_k(.)\}$'s, also known as structured random effects,
			include spline models, spatial effects, temporal effects, non-separable
			spatio-temporal effects, frailties, subject or group-specific
			intercepts and slopes etc.}
		\item {The latent field $\pmb{\mathcal{X}}$ is
			formed from the structured additive predictor as
			$\{\beta_0,\pmb{\beta},\{f_k(.)\},\pmb{\eta}\}$ with a multivariate Gaussian prior such that the latent field forms a Gaussian Markov
			random field with sparse precision matrix $\pmb{Q}(\pmb{\theta}_2)$,
			i.e.\ $\pmb{\mathcal{X}}\sim N(\pmb{0},\pmb{Q}^{-1}(\pmb{\theta}_2))$. }
		\item {A prior (possibly non-Gaussian),
			$\pmb{\pi}(\pmb{\theta})$ can then be formulated for the set of
			hyperparameters $\pmb{\theta}=(\pmb{\theta}_1,\pmb{\theta}_2)$.}
	\end{enumerate}
	From this hierarchical Bayesian formulation the
	joint posterior distribution is then given by:
	\begin{equation}
	\pmb{\pi}(\pmb{\mathcal{X}},\pmb{\theta})\propto\pmb{\pi}(\pmb{\theta})\pmb{\pi}
	(\pmb{\mathcal{X}}|\pmb{\theta})\prod_{i}\pi(y_i|\pmb{\mathcal{X}},\pmb{\theta})
	\label{postINLA}
	\end{equation}
	Within this framework the joint posterior density \eqref{postINLA} and subsequently the marginal posterior densities,
	$\pmb{\pi}(\mathcal{X}_i|\pmb{y}),i=1...n$ and
	$\pmb{\pi}(\pmb{\theta}|\pmb{y})$ can be efficiently and accurately calculated using the INLA methodology developed by \cite{rue2009}, without using Gibbs sampling.  
	\subsection{Competing risks joint models as LGMs}
	In this section we show that most competing risks joint models are indeed LGMs and can therefore be defined as one model. This enables the use of the INLA methodology for many different competing risks joint models, for example discrete competing risks joint models, spatial competing risks joint models, non-linear competing risks joint models, correlated competing risks joint models etc.\\ \\
	Consider the case with $C$ competing risks, $N$ individuals with time to event observations $\pmb{t}$ and event variable $\pmb{d}$, and $M$ series' of longitudinal observations each of size $N_l,l=1,...,M$. Let $\pmb{d}_i,i=1,...,C$ be a vector of indicators if the event type is $i$. \\ \\
	Now define the stacked response list $\pmb{y}^{\text{stacked}}=\{\pmb{y}_1, \pmb{y}_2,...,\pmb{y}_M,(\pmb{t}_1,\pmb{d}_1),(\pmb{t}_2,\pmb{d}_2),...,(\pmb{t}_C,\pmb{d}_C)\}$. Also define the stacked linear predictor $\pmb{\eta}^{\text{stacked}}=\{\pmb{\eta}^L_1, \pmb{\eta}^L_2,...,\pmb{\eta}^L_M,\pmb{\eta}^1,\pmb{\eta}^2,...,\pmb{\eta}^C,\}$ as in \eqref{longlik1} and \eqref{survlik1} where each component is of the form \eqref{additive predictor}. For each component we define two vectors of hyperparameters, $\pmb\theta_1$ for the hyperparameters from the likelihood (like a precision parameter if a Gaussian distribution is assumed, or the shape parameter for a Weibull model etc.) and $\pmb\theta_2$ for the hyperparameters from the random effects in the linear predictor (like a range parameter for a spatial effect, or a precision parameter for a spline effect etc.), such that the stacked vector of hyperparameters is $\pmb{\theta}^{\text{stacked}}=\{\pmb\theta^L_{1,1},\pmb\theta^L_{2,1},\pmb\theta^L_{1,2},\pmb\theta^L_{2,2},...,\pmb\theta^L_{1,M},\pmb\theta^L_{2,M},\pmb\theta_{1,1},\pmb\theta_{2,1},\pmb\theta_{1,2},\pmb\theta_{2,2},...,\pmb\theta_{1,C},\pmb\theta_{2,C}\}$. The latent field is then defined as $\pmb{\mathcal{X}}=\{\pmb\beta^{\text{stacked}},\pmb\eta^{\text{stacked}},\{\pmb{f}(.)\}^{\text{stacked}}\}$, and we assume a multivariate Gaussian prior for $\pmb{\mathcal{X}}$.\\ \\
	From this construction the cause-specific competing risks joint model (in general, with various random effects like spatial fields, spline effects, correlated models etc.) is an LGM and the joint posterior distribution of the latent field and the hyperparameters is 
	\begin{equation*}
		\pmb{\pi}(\pmb{\mathcal{X}},\pmb{\theta}^{\text{stacked}})\propto\pmb{\pi}(\pmb{\theta}^{\text{stacked}})\pmb{\pi}
		(\pmb{\mathcal{X}}|\pmb{\theta}^{\text{stacked}})\prod_{i=1}^{\text{length}(\pmb{y}^{\text{stacked}})}\pi(y_i|\pmb{\mathcal{X}},\pmb{\theta}^{\text{stacked}}).
	\end{equation*}
	Since this proposition is abstract we provide some specific examples of relevant competing risks joint models in the next section.
	\subsection{Specific examples}
	Next we present some specific (non-exhaustive) examples of competing risks joint models that fit in this framework.
	\subsubsection{Example 1. Poisson competing risks random-effects joint model}
	Suppose that the longitudinal observations, $y_i(t)$ follow a Poisson distribution with mean $\lambda_i(t)$, i.e., $y_i(t)\sim \text{Poisson}(\lambda_i(t))$ with $\lambda_i(t)=\exp(\eta^L_{i}(t))$.  Also assume three competing causes and a Weibull model for all the cause-specific hazard functions with shape parameters $\alpha_1,\alpha_2,\alpha_3$, i.e. for observed time $t_i$ with event indicators $\pmb{d}_i=\{d_{i1},d_{i2},d_{i3}\}$, $h_{ij}(t)=\alpha_j(\gamma_{ij})^\alpha_j t^{\alpha_j-1}$ with $\gamma_{ij}=\exp(\eta_i^{(j)}),j=1,2,3$.\\ \\
	Now define the associated linear predictors as follows:
	\begin{eqnarray}
	\eta^L(t)&=&\pmb\beta^L\pmb{X}^L(t)+\pmb{v}+\pmb{w}t\\
	\eta^{(1)}(t)&=&\pmb\beta^{(1)}\pmb{X}^{(1)}(t)+\gamma^{(1)}\pmb{v}+\kappa^{(1)}\pmb{w}t\notag\\
	\eta^{(2)}(t)&=&\pmb\beta^{(2)}\pmb{X}^{(2)}(t)+\gamma^{(2)}\pmb{v}+\kappa^{(2)}\pmb{w}t\notag\\
	\eta^{(3)}(t)&=&\pmb\beta^{(3)}\pmb{X}^{(3)}(t)+\gamma^{(3)}\pmb{v}+\kappa^{(3)}\pmb{w}t\notag,
	\label{poiscmpjoint}
	\end{eqnarray}
	where $\begin{bmatrix}
	v_i\\w_i
	\end{bmatrix}\sim N\left(\begin{bmatrix}
	0\\0
	\end{bmatrix},\begin{bmatrix}
	\tau^{-1}_v & \rho\\ \rho & \tau^{-1}_w
	\end{bmatrix}\right).$ Note that we do not enforce the same scaling parameter for the two random effects but accommodate for different scalings of the random intercept and the random slope.
	
	Now define $\pmb{\theta}=\{\alpha_1,\alpha_2,\alpha_3,\tau^{-1}_v, \tau^{-1}_w,\rho\}$ as the vector of hyperparameters. Define the latent field $$\pmb{\mathcal{X}}=\{\pmb\beta^L,\pmb\beta^{(1)},\pmb\beta^{(2)},\pmb\beta^{(3)},\pmb{\eta}^L,\pmb{\eta}^{(1)},\pmb{\eta}^{(2)},\pmb{\eta}^{(3)},\pmb{v}, \pmb{w}\},$$ 
	then this competing risks joint model is an LGM following Section \ref{seccmp}.
	\subsubsection{Example 2. Spatial competing risks joint model with smoothed longitudinal predictor as the association term and unknown non-linear time trend}
	Suppose that the longitudinal observations, $y_i$ follow a Gaussian distribution with mean $\mu_i$ and precision parameter $\tau$, i.e., $y_i\sim N(\mu_i,\tau^{-1})$ with $\mu_i=\eta^L_{i}(s,t)$ with $t$ a time index and $s$ a space index.  Also assume three competing causes and an exponential model for all the cause-specific hazard functions i.e. for observed time $t_i$ with event indicators $\pmb{d}_i=\{d_{i1},d_{i2},d_{i3}\}$, $h_{ij}(t)=\lambda_{ij}$ with $\lambda_{ij}=\exp(\eta_i^{(j)}),j=1,2,3$.\\ \\
	Now define the associated linear predictors as follows:
	\begin{eqnarray}
	\eta^L(s,t)&=&\pmb\beta^L\pmb{X}^L(t)+\pmb{v}+f(t)+u(s)\\
	\eta^{(1)}(s,t)&=&\pmb\beta^{(1)}\pmb{X}^{(1)}(t)+\gamma^{(1)}\eta^L(s,t)\notag\\
	\eta^{(2)}(s,t)&=&\pmb\beta^{(2)}\pmb{X}^{(2)}(t)+\gamma^{(2)}\eta^L(s,t)\notag\\
	\eta^{(3)}(s,t)&=&\pmb\beta^{(3)}\pmb{X}^{(3)}(t)+\gamma^{(3)}\eta^L(s,t)\notag,
	\end{eqnarray}
	where $f(t)$ is an unknown non-linear function of time (i.e. a spline component) with hyperparameters $\pmb{\theta}_{\text{time}}$ and $u(s)$ is a spatial field (eg. Besag (***) or BYM (**) models for discrete regions or SPDE (eg. Matern) (***) model for continuous space) with hyperparameters $\pmb{\theta}_{\text{space}}$.
	Now each cause-specific hazard function is associated with the longitudinal process through its own scaling $\gamma^{(j)}$.
	
	Now define $\pmb{\theta}=\{\pmb{\theta}_{\text{time}},\pmb{\theta}_{\text{space}}, \tau\}$ as the vector of hyperparameters. Define the latent field $$\pmb{\mathcal{X}}=\{\pmb\beta^L,\pmb\beta^{(1)},\pmb\beta^{(2)},\pmb\beta^{(3)},\pmb{\eta}^L,\pmb{\eta}^{(1)},\pmb{\eta}^{(2)},\pmb{\eta}^{(3)},\pmb{f}(.), \pmb{u}(.)\},$$ 
	then this spatial competing risks joint model is an LGM following Section \ref{seccmp}.
	 
	 	\section{Application - SANAD trial}\label{SANAD}
\begin{figure}[h]
	 	\begin{center}
	 		\includegraphics[width=12cm]{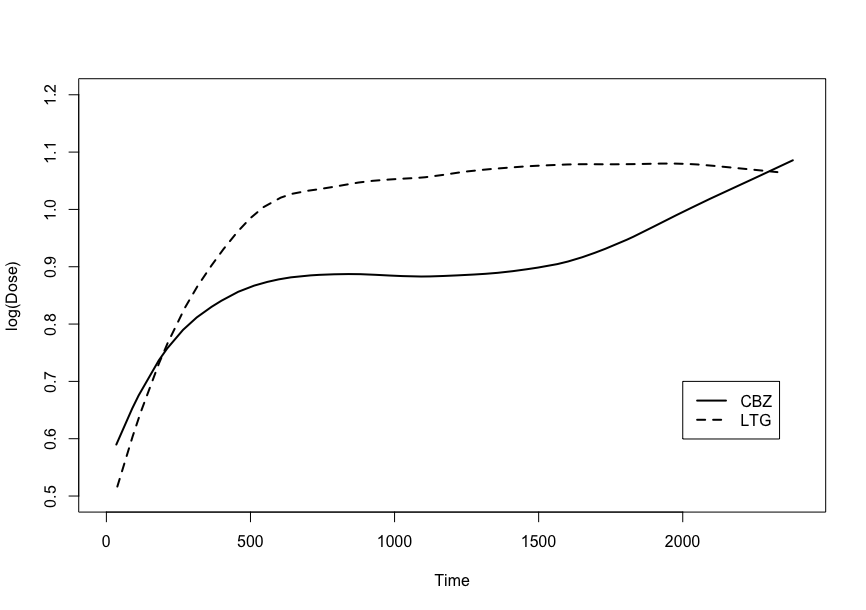}
	 	\end{center}
	 	\label{meanplotapp}
	 	\caption{Mean trajectories for CBZ and LTG treatment groups}
	 \end{figure}

	 The SANAD trial \citep{marson2007} was a randomized trial study to investigate the effect of seizure control drugs in patient with epilepsy for whom carbamazepine (CBZ) was considered the standard treatment protocol. The patients were randomly assigned to CBZ, lamotrigine (LTG), gabapentin, oxcarbazepine and topiramate. In the main analysis, LTG appeared to be superior to CBZ but since then various criticisms regarding this finding have risen, mainly due to the difference in titration speed and therefore the dosage. In \cite{williamson2008} this data was analyzed using a competing risks random-effects joint model with a changepoint at time $t=500$. In Figure \ref{meanplotapp} the mean trajectories for the two treatment groups CBZ and LTG are presented. It is clear that the mean trajectories are firstly different between the two treatments and secondly, the changepoint at $t=500$ seems quite subjective since for CBZ the trajectory does not seems to have a constant linear trend after $t=500$. We therefore include a spline component in the form of a random walk order two model for the dosage trajectory over time. Also, we do not enforce equal scaling for the random intercepts and slopes. 
	 
	 \subsection{Bayesian competing risks joint model}
	 We propose the following competing risks joint model with non-linear longitudinal trajectories for the SANAD trial data,
	 \begin{eqnarray}
	 \eta^L(t)&=&\beta_0+\beta_1 \text{CBZ}+\text{CBZ}f_{\text{CBZ}}(t)+\text{LTG}f_{\text{LTG}}(t)\\
	 \eta^{\text{(UAE)}}(s)&=&\beta_0+\beta^{(1)}_1 \text{CBZ}+\gamma_1\eta^L(s)\notag\\
	 \eta^{(\text{ISC)}}(s)&=&\beta_0+\beta^{(2)}_1 \text{CBZ}+\gamma_2\eta^L(s)\notag,
	 \end{eqnarray}
	 where $f_{\text{CBZ}}$ is a random walk order two model \citep{lindgren2008} with hyperparameter $\tau_{f_{\text{CBZ}}}$ on the interaction of the time and $\text{CBZ}$, the CBZ treatment indicator, and similarly for LTG. We assume a Gaussian longitudinal distribution with noise precision $\tau_L$, and Weibull models for the cause-specific hazard functions with hyperparameters $\alpha_{\text{UAE}},\alpha_\text{ISC}$.
	 The priors for the latent field are all Gaussian i.e.,
	 \begin{eqnarray}
	 \beta_0\sim N(0,\tau_{\beta_0}^{-1}),\quad
	 \gamma_1\sim N(0,\tau_{\gamma_1}^{-1}),\quad
	 \gamma_2\sim N(0,\tau_{\gamma_2}^{-1}),\notag
	 \end{eqnarray}
	 with penalizing complexity priors \citep{simpson2017} for most of the hyperparameters, such that
	 \begin{eqnarray}
	 &&\tau^{-0.5}_L\sim \text{Exp}(\lambda_L), \quad
	 \tau_{\beta_0}^{-0.5}\sim \text{Exp}(\lambda_{\beta_0}), \quad
	 \tau_{\gamma_1}^{-0.5}\sim \text{Exp}(\lambda_{\gamma_1}),\notag\\
	 &&\tau_{\gamma_2}^{-0.5}\sim \text{Exp}(\lambda_{\gamma_2}),\quad
	 \tau_{f_{\text{CBZ}}}^{-0.5}\sim \text{Exp}(\lambda_\text{CBZ}),\quad
	 \tau_{f_{\text{LTG}}}^{-0.5}\sim \text{Exp}(\lambda_\text{CBZ}),\notag\\
	 &&10\ln(\alpha_{\text{UAE}})\sim N(0,\tau_\text{UAE}^{-1}), \quad 10\ln(\alpha_{\text{ISC}})\sim N(0,\tau_\text{ISC}^{-1})\notag
	 \end{eqnarray}
	 such that the user-defined parameter $\lambda_{[.]}$ is defined through the probability
	 $P(\tau^{-0.5}_{[.]}>1)$, as an example $P(\tau^{-0.5}_{[.]}>1)=0.01$ produces $\lambda_{[.]}=\ln(0.01)$. Note that the exponential prior on $\sigma_{[.]}=\tau^{-0.5}_{[.]}$ translates to a Gumbel type 2 prior on $\tau_{[.]}$.
	 The motivation for employing
	 penalizing complexity priors for the precision hyperparameters are
	 founded in the fact that the usual priors for the variance components,
	 i.e. independent inverse-gamma priors as in \cite{huang2018}, overfits
	 in the sense that it cannot contract to the simpler model in which the respective model
	 component has trivial variance. This is especially important in the
	 case of joint models since the effect of overfitting is exacerbated by
	 the influence of the shared random effect on all the linear
	 predictors.
	 \subsection{Results}
\begin{figure}[h]
	 	\begin{center}
	 		\includegraphics[width=12cm]{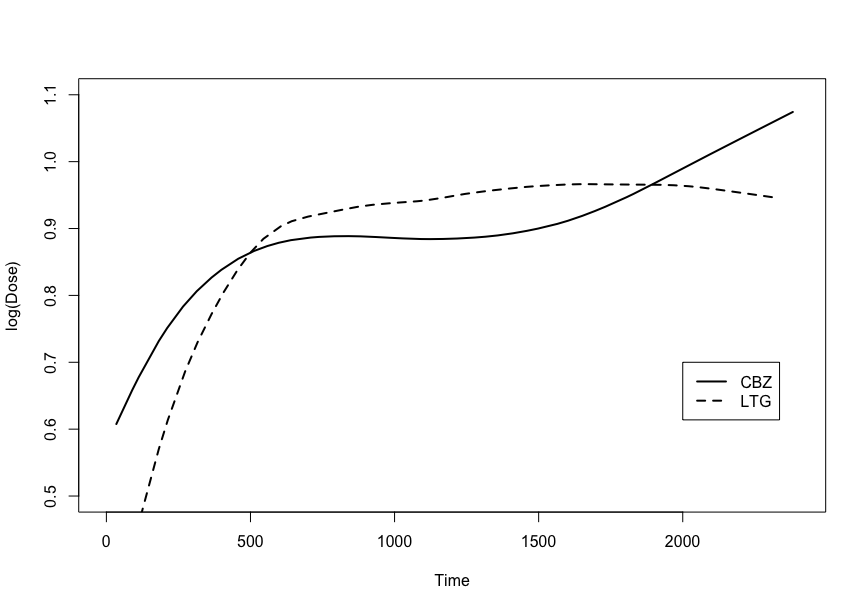}
	 	\end{center}
	 	\label{figmodel}
	 	\caption{Mean estimated dosage trajectories for the CBZ and LTG treatment groups}
	 \end{figure}

	 The estimated mean dosage trajectories for the two treatements, CBZ and LTG, on a log scale is presented in Figure \ref{figmodel}. The effect of the joint model on the longitudinal trajectories is clear. The dropout process affects the longitudinal trajectories in a meaningful way.
	 \\
	 The estimated competing risks joint model is:
	 \begin{eqnarray}
	 \widehat{\eta}^L(t)&=&2.438(2.319;2.556)+0.305(0.145;0.465) \text{CBZ}+\text{CBZ}f_{\text{CBZ}}(t)+\text{LTG}f_{\text{LTG}}(t)\notag\\
	 \widehat{\eta}^{\text{(UAE)}}(s)&=&2.438(2.319;2.556)-0.726(-1.104;-0.371) \text{CBZ}-1.1(-1.192;-1.02)\eta^L(s)\notag\\
	 \widehat{\eta}^{(\text{ISC)}}(s)&=&2.438(2.319;2.556)-0.11(-0.406;0.175) \text{CBZ}-0.998(-1.087;-0.928)\eta^L(s)\notag,
	 \end{eqnarray}
	 with $\tau_{f_{\text{CBZ}}}=14.879$, $\tau_{f_{\text{LTG}}}=7.055$, $\alpha_{\text{UAE}}=0.84(0.05;0.915)$ and $\alpha_{\text{ISC}}=1.21(1.19;1.33)$. \\ \\
	 The negative association between the cause-specific hazard  functions of UAE and ISC, repsectively, and the smoothed dosage level should be interpreted with the shape parameters, $\alpha_{\text{UAE}}$ and $\alpha_{\text{ISC}}$ of the weibull models. For UAE, the estimated shape parameter is significantly smaller than $1$ which implies that as the dose increase over time as captured by $\widehat{\eta}^L(t)$ the linear predictor $\widehat{\eta}^{\text{(UAE)}}(s)$ will decrease due to the negative association and even more so for treatment CBZ. The decreased $\widehat{\eta}^{\text{(UAE)}}(s)$ together with $\alpha_{\text{UAE}}=0.84$ leads to the conclusion that the hazard of UAE is lower for LTG than for CBZ, and increasing over time since higher doses correspond to larger times. The converse is true for the hazard of ISC since $\alpha_{\text{ISC}}$ is significantly larger than $1$. Due to the insignificant effect of treatment on the hazard of ISC, the two treatments under consideration provide similar seizure control.  \\ \\
	 Our findings are thus consistent with those of \citet{williamson2008} with the added advantage of the flexible estimated mean longitudinal trajectories for the two treatment dosages.
	
	\section{Simulation study - Discrete competing risks random-effects joint model}\label{simstudy}
\begin{figure}[h]
			\begin{center}
		\includegraphics[width=12cm]{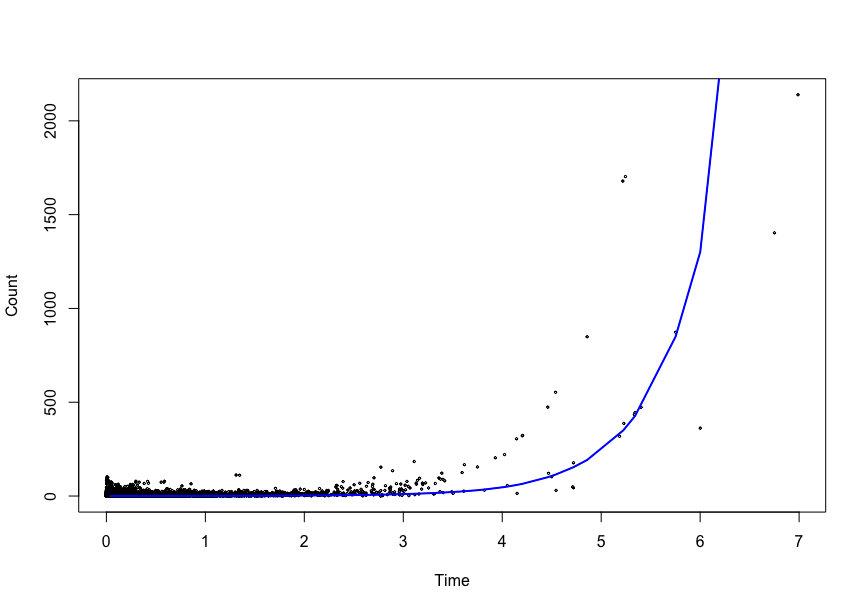}
	\end{center}
		\caption{Simulated longitudinal data and estimated mean trajectory}
		\label{figsim}
	\end{figure}

	In this section we simulate longitudinal count data as well as time to event data with three competing risks (Causes $1,2$ and $3$). We also included a non-linear time component in the longitudinal trajectory. The data is simulated from a competing risks joint random-effects model with a random intercept (similar to \eqref{poiscmpjoint}), such that for $N$ individuals, each with $N_i,i=1,...,N$ longitudinal counts and time to event $t_i$ with event indicator $d_i\in\{0,1,2,3\}$ the model is
	\begin{eqnarray}
	y_i(t)&\sim&\text{Poisson}(\exp(t^{1.2}+u_i))\notag\\
	h_1(t)&=&\exp(0.3u_i+0.01\text{Age})\notag\\
	h_2(t)&=&\exp(-0.1u_i+0.015\text{Age})\notag\\
	h_3(t)&=&\exp(0.2u_i+0.0003\text{Age}),\notag
    \end{eqnarray}
	with Age a discrete explanatory variable within the range $(15;75)$ and $u_i\sim N(0,1)$ is a random intercept which associates the longitudinal series with the cause-specific hazard functions, $h_j(t),j=1,2,3$. Note that we also include a non-linear trend in the longitudinal trajectory, $t^{1.2}$. In the simulation setup, we induce a negative association between the hazard of event $2$ and the longitudional series, whereas the hazard of events $1$ and $3$ are positively correlated with the longitudinal series. From the above, the hazard functions are constant over time and an exponential distribution is thus implied.\\ \\ In Figure \ref{figsim} the longitudinal series and the mean estimated trajectory from a random walk order two model with precision $\tau_r$ is presented. 
The results are presented in Table \ref{simtable} and it is clear that this challenging model is recovered well.
\begin{table}[h]
	\begin{center}
	\begin{tabular}{|c||c|c|c|}
		\hline
		\textbf{Parameter} & \textbf{True value} & \textbf{Estimate} & $95\%$ \textbf{credible interval}\\ \hline
		$\sigma_u$ & $1$ & $1.017$ & $(0.925;1.114)$\\ \hline
		$\beta_1$ & $0.01$ & $0.012$ &$(0.005;0.018)$\\ \hline
		$\beta_2$ & $0.015$ & $0.052$& $(0.009;0.019)$\\ \hline
		$\beta_3$ & $0.0003$ & $0.0001$ &$(-0.003;0.005)$\\ \hline
		$\gamma_1$ & $0.3$ & $0.285$ &$(0.192;0.369)$\\ \hline
		$\gamma_2$ & $-0.1$ & $-0.108$& $(-0.206;-0.011)$\\ \hline
		$\gamma_3$ & $0.2$ & $0.208$& $(0.062;0.352)$\\ \hline
		$\tau_r$ & $.$ & $28.513$ & $(11.709;116.878)$\\  \hline
	\end{tabular}
\end{center}
\caption{Estimates from the discrete competing risks joint model using R-INLA}
\label{simtable}
\end{table} 
\newpage

	\section{Discussion}\label{disc}
	Here we presented a general method to apply a wide class of competing risks joint models using the R-INLA framework. The main stimulus for this research was the necessity for a proper non-linear competing risks joint model as motivated from the SANAD trial. Although we addressed this issue, we also developed a class of latent Gaussian competing risks joint models which enables the practitioner to use our method for various realistic situations which demands different types of competing risks joint models. We presented the details of two specific realistic situations, one being a competing risks model with three competing events and one longitudinal series of count data, the other a spatial competing risks joint model. The application of the non-linear competing risks joint model to the SANAD trial proved fruitful as this model properly captures the non-linear trajectory. \\
 The possibility of a non-linear competing risks joint model based on a random walk order two model emerged through the R-INLA framework. The INLA methodology \citep{rue2009} has been proved to be a very efficient and popular computational platform for Bayesian models. The use of the R-INLA framework necessitated the formulation of the proposed model as a latent Gaussian model, which in turn facilitated the formulation of various other competing risks joint models that could arise in practical situations, such as multiple longitudinal series, discrete longitudinal distributions, spatial models and frailty models to name but a few.\\ \\ 
 The possibilities of realistic competing risks joint models that can be inferred through the proposed methodology is vast and exciting. We only presented a selected few which we deemed interesting and applicable. A thorough presentation of all the possibilities is not possible in this paper but we believe that the general method presented herein catalyzes future research, especially different instances of competing risks joint models motivated from a modeling perspective.
	\section*{Code appendix}
	\begin{lstlisting}
		
		#Competing risks simulated example
		N=1000 #number of patients
		N_i=round(runif(N,min=10,max=15)) #Number of longitudinal observations per 
		person
		#Generate time-dependent exponential event times with iid shared random 
		effects term
		#Survival component
		u_i=1+rnorm(N,mean=0,sd=1) #iid random effect
		coeff=0.5 #share parameter
		lambda=exp(coeff*u_i)
		s_i=rexp(N,rate=lambda) #exponential survival times
		#library(purrr)
		#c_i=rbernoulli(N,p=0.9)#censoring
		c_i=rep(1,N) #no censoring
		age=round(runif(N,15,75))
		eta1=0.3*u_i+0.01*age
		eta2=-0.1*u_i+0.02*age
		eta3=0.2*u_i+0.0003*age
		lambda1=exp(eta1)
		lambda2=exp(eta2)
		lambda3=exp(eta3)
		
		IDs=1:N
		time1=rexp(n=N,rate=lambda1)
		Cause=rbinom(n=N,3,0.6)
		time=rep(NA,N)
		for (i in 1:N)
		{if (Cause[i]==1) {time[i]=rexp(n=1,rate=lambda1[i])}
			if (Cause[i]==2) {time[i]=rexp(n=1,rate=lambda2[i])}
			if (Cause[i]==3) {time[i]=rexp(n=1,rate=lambda3[i])}
			if (Cause[i]==0) {time[i]=rexp(n=1,rate=1)}
		}
		
		dataS<-data.frame(ID=IDs,C=Cause,Time=time,Age=age)
		data1<-dataS
		dataE1<-data1
		dataE1$event<-dataE1$C
		dataE1$event[dataE1$event!=1]<-0
		dataE2<-data1
		dataE2$event<-dataE2$C
		dataE2$event[dataE2$event!=2]<-0
		dataE2$event<-dataE2$event/2
		dataE3<-data1
		dataE3$event<-dataE3$C
		dataE3$event[dataE3$event!=3]<-0
		dataE3$event<-dataE3$event/3
		
		#Longitudinal component
		t=rep(NA,sum(N_i))
		ID=rep(NA,sum(N_i))
		a=rep(NA,sum(N_i))
		age_i=rep(NA,sum(N_i))
		t[1:N_i[1]]=runif(N_i[1],min=0,max=s_i[1]) # observation times 
		ID[1:N_i[1]]=1
		a[1:N_i[1]]=u_i[1]
		age_i[1:N_i[1]]=age[1]
		
		for (i in 2:N){
			t[(sum(N_i[1:(i-1)])+1):(sum(N_i[1:(i-1)])+N_i[i])]=runif(N_i[i],
			min=0,max=s_i[i])
			 #observation times
			ID[(sum(N_i[1:(i-1)])+1):(sum(N_i[1:(i-1)])+N_i[i])]=i #person ID
			a[(sum(N_i[1:(i-1)])+1):(sum(N_i[1:(i-1)])+N_i[i])]=u_i[i] 
			#iid random effect
			age_i[(sum(N_i[1:(i-1)])+1):(sum(N_i[1:(i-1)])+N_i[i])]=age[i] 
			#iid random effect
		}
		y=rpois(n=sum(N_i),lambda=(exp(t^(1.2)+a))) #longitudinal response with N_i[i]
		 observations per person and a t^2 trend
		plot(t,y,pch=1,cex=0.3) #view longitudional data
		
		#INLA on sim data
		dataL<-data.frame(Time=t,Age=age_i,Y=y,ID=ID)
		nL<-nrow(dataL)
		nS<-nrow(dataS)
		
		fixed.eff<-data.frame(mu=as.factor(c(rep(1,nL),rep(2,nS),rep(3,nS),
		rep(4,nS))),
		ageL=c(dataL$Age,rep(0,nS),rep(0,nS),rep(0,nS)),
		age1=c(rep(0,nL),dataS$Age,rep(0,nS),rep(0,nS)),
		age2=c(rep(0,nL),rep(0,nS),dataS$Age,rep(0,nS)),
		age3=c(rep(0,nL),rep(0,nS),rep(0,nS),dataS$Age),
		Ltime=c(dataL$Time,rep(0,nS),rep(0,nS),rep(0,nS)),
		C1time=c(rep(0,nL),dataS$Time,rep(0,nS),rep(0,nS)),
		C2time=c(rep(0,nL),rep(0,nS),dataS$Time,rep(0,nS)),
		C3time=c(rep(0,nL),rep(0,nS),rep(0,nS),dataS$Time))
		random.eff<-list(timeL=c(dataL$Time,rep(NA,nS),rep(NA,nS),rep(NA,nS)),
		Lr1=c(dataL$ID,rep(NA,nS),rep(NA,nS),rep(NA,nS)),
		Lr2=c(dataL$ID,rep(NA,nS),rep(NA,nS),rep(NA,nS)),
		C1r1=c(rep(NA,nL),dataS$ID,rep(NA,nS),rep(NA,nS)),
		C1r2=c(rep(NA,nL),dataS$ID,rep(NA,nS),rep(NA,nS)),
		C2r1=c(rep(NA,nL),rep(NA,nS),dataS$ID,rep(NA,nS)),
		C2r2=c(rep(NA,nL),rep(NA,nS),dataS$ID,rep(NA,nS)),
		C3r1=c(rep(NA,nL),rep(NA,nS),rep(NA,nS),dataS$ID),
		C3r2=c(rep(NA,nL),rep(NA,nS),rep(NA,nS),dataS$ID))
		
		
		jointdata<-c(fixed.eff,random.eff)
		y.long <- c(dataL$Y,rep(NA, nS),rep(NA,nS),rep(NA,nS))
		y.survC1 <- inla.surv(time = c(rep(NA, nL),dataS$Time,rep(NA,nS),rep(NA,nS)), 
		event = c(rep(NA, nL),dataE1$event,rep(NA,nS),rep(NA,nS)))
		y.survC2 <- inla.surv(time = c(rep(NA, nL), rep(NA,nS),dataS$Time, 
		rep(NA,nS)),
		 event = c(rep(NA, nL),rep(NA,nS),dataE2$event,rep(NA,nS)))
		y.survC3 <- inla.surv(time = c(rep(NA, nL), rep(NA,nS), rep(NA,nS),
		dataS$Time),
		 event = c(rep(NA, nL),rep(NA,nS),rep(NA,nS),dataE3$event))
		
		y.joint<-list(y.long,y.survC1,y.survC2,y.survC3)
		
		jointdata$Y=y.joint
		
		#Model fit
		formula.model=Y~-1+mu+age1+age2+age3+
		f(inla.group(timeL,n=50),model="rw2",scale.model = TRUE,hyper = list(prec =
		 list(prior="pc.prec", param=c(1, 0.01))))+
		f(Lr1, model="iid")+ 
		f(C1r1, copy="Lr1",fixed=FALSE)+
		f(C2r1, copy="Lr1",fixed=FALSE)+
		f(C3r1, copy="Lr1",fixed=FALSE)
		
		Jointmodel= inla(formula.model, family = c("poisson","exponentialsurv","exponentialsurv",
		"exponentialsurv"),
		data = jointdata, verbose=TRUE)
		
		summary(Jointmodel)
		
		plot(t,y,pch=1,cex=0.3,ylab="Count",xlab="Time") #view longitudional data
		lines(Jointmodel$summary.random$`inla.group(timeL, n = 50)`[,1],exp(Jointmodel$summary.random$`inla.group(timeL, n = 50)`[,2]),
		col="blue",lwd=2)
		
		##########################################################
		#SANAD - epileptic
		library(joineR)
		library(INLA)
		
		mTime=max(max(epileptic$time),max(epileptic$with.time))
		epileptic$time<-epileptic$time/mTime
		epileptic$with.time <-epileptic$with.time/mTime
		
		epileptic$interaction <- with(epileptic, time * (treat == "LTG"))
		epileptic$interaction2 <- with(epileptic, time * (treat == "CBZ"))
		epileptic$interaction[epileptic$interaction==0]<-NA
		epileptic$interaction2[epileptic$interaction2==0]<-NA
		
		data1<-epileptic
		dataL<-data.frame(ID=data1$id,LDose=log(data1$dose+0.1),Time=data1$time,
		TimeSp=data1$time,Age=data1$age,Gender=data1$gender,LD=data1$learn.dis,
		Treatment=data1$treat,InteractionLTG=data1$interaction,InteractionCBZ=
		data1$interaction2,list(V=data1$id,W=data1$id),Dose=data1$dose)
		dataS<-data.frame(ID=data1$id,Time=as.numeric(data1$with.time),Status=
		data1$with.status2,StatusISC=data1$with.status.isc,StatusUAE=
		data1$with.status.uae,Age=data1$age,Gender=data1$gender,LD=data1$learn.dis,
		Treatment=data1$treat)
		dataS<-subset(dataS,!duplicated(dataS$ID))
		summary(dataL)
		summary(dataS)
		
		#Visualize
		plot(dataL$Time[dataL$ID==42],dataL$Dose[dataL$ID==42],type="n",
		xlim=c(0,1),ylim=c(0,10))
		for (i in unique(dataL$ID)){
			lines(dataL$Time[dataL$ID==i],dataL$Dose[dataL$ID==i],col=i)
		}
		
		dataLCBZ=dataL[dataL$Treatment=="CBZ",]
		dataLLTG=dataL[dataL$Treatment=="LTG",]
		
		#Mean trajectories
		modCBZ<-inla(formula=dataLCBZ$Dose~f(inla.group(dataLCBZ$Time,n=50),
		model="rw2",scale.model = TRUE,hyper = list(prec = list(prior="pc.prec", 
		param=c(1, 0.01)))),data=dataLCBZ,family="gaussian",control.compute=
		list(dic=TRUE,config=TRUE))
		modLTG<-inla(formula=dataLLTG$Dose~f(inla.group(dataLLTG$Time,n=50),
		model="rw2",scale.model = TRUE,hyper = list(prec = list(prior="pc.prec", 
		param=c(1, 0.01)))),data=dataLLTG,family="gaussian",control.compute=
		list(dic=TRUE,config=TRUE))
		#Plot means
		plot(modCBZ$summary.random$`inla.group(dataLCBZ|S|Time, n =
		 50)`[,1]*2400,log(modCBZ$summary.fixed[1,1]+
		modCBZ$summary.random$`inla.group(dataLCBZ|S|Time, n = 50)`[,2]),
		type="l",lwd=2,ylim=c(0.5,1.2),xlab="Time",ylab="log(Dose)",
		xlim=c(0,2400))
		lines(modLTG$summary.random$`inla.group(dataLLTG|S|Time, n = 50)`[,1]
		*2400,log(modLTG$summary.fixed[1,1]+
		modLTG$summary.random$`inla.group(dataLLTG|S|Time, n = 50)`[,2]),
		lty=2,lwd=2)
		legend(x=2000,y=0.7,legend=c("CBZ","LTG"),lty=c(1,2),lwd=c(2,2))
		
		#Joint model
		#Pre-work for entire predictor
		nL<-nrow(dataL)
		nS<-nrow(dataS)
		fixed.eff<-data.frame(mu=as.factor(c(rep(1,nL),rep(1,nL),rep(2,nS),
		rep(3,nS))),
		ageL=c(dataL$Age,dataL$Age,rep(0,nS),rep(0,nS)),
		ageUAE=c(rep(0,nL),rep(0,nL),dataS$Age,rep(0,nS)),
		ageISC=c(rep(0,nL),rep(0,nL),rep(0,nS),dataS$Age),
		treatmentL=as.factor(c(dataL$Treatment,dataL$Treatment,rep(NA,nS),
		rep(NA,nS))),
		treatmentUAE=as.factor(c(rep(NA,nL),rep(NA,nL),dataS$Treatment,
		rep(NA,nS))),
		treatmentISC=as.factor(c(rep(NA,nL),rep(NA,nL),rep(NA,nS),
		dataS$Treatment)),
		genderL=as.factor(c(dataL$Gender ,dataL$Gender ,rep(NA,nS),
		rep(NA,nS))),
		genderUAE=as.factor(c(rep(NA,nL),rep(NA,nL),dataS$Gender ,
		rep(NA,nS))),
		genderISC=as.factor(c(rep(NA,nL),rep(NA,nL),rep(NA,nS),
		dataS$Gender)))
		random.eff<-list(timeL_LTG=c(dataL$InteractionLTG,rep(NA,nL),
		rep(NA,nS),rep(NA,nS)),
		timeL_CBZ=c(dataL$InteractionCBZ,rep(NA,nL),rep(NA,nS),
		rep(NA,nS)),
		linpredL=c(rep(NA,nL),dataL$ID,rep(NA,nS),rep(NA,nS)),
		linpredL2=c(rep(NA,nL),rep(-1,nL),rep(NA,nS),rep(NA,nS)),
		betaUAE=c(rep(NA,nL),rep(NA,nL),dataS$ID,rep(NA,nS)),
		betaISC=c(rep(NA,nL),rep(NA,nL),rep(NA,nS),dataS$ID),
		frailtyUAE=c(rep(NA,nL),rep(NA,nL),dataS$ID,rep(NA,nS)),
		frailtyISC=c(rep(NA,nL),rep(NA,nL),rep(NA,nS),dataS$ID))
		
		
		jointdata<-c(fixed.eff,random.eff)
		y.long <- c(dataL$Dose,rep(NA,nL),rep(NA, nS),rep(NA,nS))
		y.eta<-c(rep(NA,nL),rep(0,nL),rep(NA,nS),rep(NA,nS))
		y.survUAE <- inla.surv(time = c(rep(NA, nL), rep(NA,nL),
		dataS$Time,rep(NA,nS)),
		 event = c(rep(NA, nL),rep(NA,nL),dataS$StatusUAE,rep(NA,nS)))
		y.survISC <- inla.surv(time = c(rep(NA, nL), rep(NA,nL),
		rep(NA,nS),dataS$Time),
		 event = c(rep(NA, nL),rep(NA,nL),rep(NA,nS),dataS$StatusISC))
		y.joint<-list(y.long,y.eta,y.survUAE,y.survISC)
		
		jointdata$Y=y.joint
		
		#Model fit
		formula.model=Y~treatmentL+treatmentUAE+treatmentISC+
		f(inla.group(timeL_LTG,n=50),model="rw2",scale.model = TRUE,hyper 
		= list(prec = list(prior="pc.prec", param=c(1, 0.01))))+
		f(inla.group(timeL_CBZ,n=50),model="rw2",scale.model = TRUE,hyper 
		= list(prec = list(prior="pc.prec", param=c(1, 0.01))))+
		f(linpredL, linpredL2, model="iid", hyper = list(prec = list(initial = -6,
		 fixed=TRUE))) + 
		f(betaUAE, copy="linpredL", hyper = list(beta = list(fixed = FALSE)))+
		f(betaISC, copy="linpredL", hyper = list(beta = list(fixed = FALSE)))
		
		
		Jointmodel= inla(formula.model, family = c("gaussian","gaussian",
		"weibullsurv","weibullsurv"),
		data = jointdata, verbose=TRUE, control.compute=list(dic=TRUE),
		control.family = list(
		list(),
		list(hyper = list(prec = list(initial = 10, fixed=TRUE))),
		list(),
		list()
		)
		)
		
		summary(Jointmodel)
		
		##Random effects only
		fixed.eff<-data.frame(mu=as.factor(c(rep(1,nL),rep(2,nS),rep(3,nS))),
		ageL=c(dataL$Age,rep(0,nS),rep(0,nS)),
		ageUAE=c(rep(0,nL),dataS$Age,rep(0,nS)),
		ageISC=c(rep(0,nL),rep(0,nS),dataS$Age),
		treatmentL=as.factor(c(dataL$Treatment,rep(NA,nS),rep(NA,nS))),
		treatmentUAE=as.factor(c(rep(NA,nL),dataS$Treatment,rep(NA,nS))),
		treatmentISC=as.factor(c(rep(NA,nL),rep(NA,nS),dataS$Treatment)),
		genderL=as.factor(c(dataL$Gender ,rep(NA,nS),rep(NA,nS))),
		genderUAE=as.factor(c(rep(NA,nL),dataS$Gender ,rep(NA,nS))),
		genderISC=as.factor(c(rep(NA,nL),rep(NA,nS),dataS$Gender)),
		UAETime=c(rep(0,nL),dataS$Time,rep(0,nS)),
		ISCTime=c(rep(0,nL),rep(0,nS),dataS$Time),
		Ltime=c(dataL$Time,rep(0,nS),rep(0,nS)))
		random.eff<-list(timeL=c(dataL$Time,rep(NA,nS),rep(NA,nS)),
		Lr1=c(dataL$ID,rep(NA,nS),rep(NA,nS)),
		Lr2=c(dataL$ID,rep(NA,nS),rep(NA,nS)),
		UAEr1=c(rep(NA,nL),dataS$ID,rep(NA,nS)),
		UAEr2=c(rep(NA,nL),dataS$ID,rep(NA,nS)),
		ISCr1=c(rep(NA,nL),rep(NA,nS),dataS$ID),
		ISCr2=c(rep(NA,nL),rep(NA,nS),dataS$ID),
		frailtyUAE=c(rep(NA,nL),dataS$ID,rep(NA,nS)),
		frailtyISC=c(rep(NA,nL),rep(NA,nS),dataS$ID))
		
		
		jointdata<-c(fixed.eff,random.eff)
		y.long <- c(dataL$Dose,rep(NA, nS),rep(NA,nS))
		y.survUAE <- inla.surv(time = c(rep(NA, nL),dataS$Time,rep(NA,nS)), 
		event = c(rep(NA, nL),dataS$StatusUAE,rep(NA,nS)))
		y.survISC <- inla.surv(time = c(rep(NA, nL), rep(NA,nS),dataS$Time),
		event = c(rep(NA, nL),rep(NA,nS),dataS$StatusISC))
		y.joint<-list(y.long,y.survUAE,y.survISC)
		
		jointdata$Y=y.joint
		
		#Model fit
		formula.model=Y~treatmentL+treatmentUAE+treatmentISC+
		f(Lr1, model="iid2d",n=2*(nL+nS+nS)) + 
		f(Lr2, Ltime,copy="Lr1",fixed=TRUE)+
		f(UAEr1,copy="Lr1",fixed=FALSE)+f(UAEr2,UAETime,copy="UAEr1",fixed=FALSE)+
		f(ISCr1,copy="Lr1",fixed=FALSE)+f(ISCr2,ISCTime,copy="ISCr1",fixed=FALSE)
		
		Jointmodel= inla(formula.model, family = c("gaussian","weibullsurv",
		"weibullsurv"),
		data = jointdata, verbose=TRUE, control.compute=list(dic=TRUE))
		
		summary(Jointmodel)
		\end{lstlisting}
	\bibliography{BioJ}

\begin{thebibliography}{}

\bibitem[Altshuler, 1970]{altshuler1970}
Altshuler, B. (1970).
\newblock Theory for the measurement of competing risks in animal experiments.
\newblock {\em Mathematical Biosciences}, 6:1--11.

\bibitem[{\'A}lvaro-Meca et~al., 2013]{alvaro2013}
{\'A}lvaro-Meca, A., Akerkar, R., Alvarez-Bartolome, M., Gil-Prieto, R., Rue,
  H., and de~Miguel, {\'A}.~G. (2013).
\newblock Factors involved in health-related transitions after curative
  resection for pancreatic cancer. 10-years experience: a multi state model.
\newblock {\em Cancer epidemiology}, 37(1):91--96.

\bibitem[Bakka et~al., 2018]{bakka2018}
Bakka, H., Rue, H., Fuglstad, G.-A., Riebler, A., Bolin, D., Illian, J.,
  Krainski, E., Simpson, D., and Lindgren, F. (2018).
\newblock Spatial modeling with {R-INLA}: A review.
\newblock {\em Wiley Interdisciplinary Reviews: Computational Statistics},
  10(6):e1443.

\bibitem[Elashoff et~al., 2007]{elashoff2007}
Elashoff, R.~M., Li, G., and Li, N. (2007).
\newblock An approach to joint analysis of longitudinal measurements and
  competing risks failure time data.
\newblock {\em Statistics in Medicine}, 26(14):2813--2835.

\bibitem[Fine and Gray, 1999]{fine1999}
Fine, J.~P. and Gray, R.~J. (1999).
\newblock A proportional hazards model for the subdistribution of a competing
  risk.
\newblock {\em Journal of the American statistical association},
  94(446):496--509.

\bibitem[Gray, 2014]{cmprskpkg}
Gray, B. (2014).
\newblock {\em cmprsk: Subdistribution Analysis of Competing Risks}.

\bibitem[Hoel, 1972]{hoel1972}
Hoel, D.~G. (1972).
\newblock A representation of mortality data by competing risks.
\newblock {\em Biometrics}, pages 475--488.

\bibitem[Hu et~al., 2009]{hu2009}
Hu, W., Li, G., and Li, N. (2009).
\newblock A bayesian approach to joint analysis of longitudinal measurements
  and competing risks failure time data.
\newblock {\em Statistics in medicine}, 28(11):1601--1619.

\bibitem[Huang et~al., 2018]{huang2018}
Huang, Y., Lu, X., Chen, J., Liang, J., and Zangmeister, M. (2018).
\newblock Joint model-based clustering of nonlinear longitudinal trajectories
  and associated time-to-event data analysis, linked by latent class
  membership: with application to {AIDS} clinical studies.
\newblock {\em Lifetime data analysis}, 24(4):699--718.

\bibitem[Krainski et~al., 2018]{krainski2018}
Krainski, E.~T., G{\'o}mez-Rubio, V., Bakka, H., Lenzi, A., Castro-Camilo, D.,
  Simpson, D., Lindgren, F., and Rue, H. (2018).
\newblock {\em Advanced spatial modeling with stochastic partial differential
  equations using R and INLA}.
\newblock Chapman and Hall/CRC.

\bibitem[Lindgren and Rue, 2008]{lindgren2008}
Lindgren, F. and Rue, H. (2008).
\newblock On the second-order model for irregular locations.
\newblock {\em Scandinavian Journal of Statistics}, 35(4):691--700.

\bibitem[Marson et~al., 2007]{marson2007}
Marson, A., Al-Kharusi, A., Alwaidh, M., Appleton, R., Baker, G., Chadwick, D.,
  Cramp, C., Cockerell, O., Cooper, P., Doughty, J., and Eaton, B. (2007).
\newblock Carbamazepine versus gabapentin, lamotrigine, oxcarbazepine or
  topiramate for partial epilepsy: results from arm a of the sanad trial.
\newblock {\em The Lancet}, 369:1000--1015.

\bibitem[Martino et~al., 2011]{martino2011}
Martino, S., Akerkar, R., and Rue, H. (2011).
\newblock Approximate {Bayesian} inference for survival models.
\newblock {\em Scandinavian Journal of Statistics}, 38(3):514--528.

\bibitem[Moeschberger and David, 1971]{moeschberger1971}
Moeschberger, M. and David, H. (1971).
\newblock Life tests under competing causes of failure and the theory of
  competing risks.
\newblock {\em Biometrics}, pages 909--933.

\bibitem[Philipson et~al., 2018]{joinerpkg}
Philipson, P., Sousa, I., Diggle, P.~J., Williamson, P., Kolamunnage-Dona, R.,
  Henderson, R., Hickey, G.~L., and Sudell, M. (2018).
\newblock {\em joineR: Joint Modelling of Repeated Measurements and
  Time-to-Event Data}.

\bibitem[Prentice et~al., 1978]{prentice1978}
Prentice, R.~L., Kalbfleisch, J.~D., Peterson~Jr, A.~V., Flournoy, N.,
  Farewell, V.~T., and Breslow, N.~E. (1978).
\newblock The analysis of failure times in the presence of competing risks.
\newblock {\em Biometrics}, pages 541--554.

\bibitem[Rue et~al., 2009]{rue2009}
Rue, H., Martino, S., and Chopin, N. (2009).
\newblock Approximate {Bayesian} inference for latent {Gaussian} models by
  using integrated nested laplace approximations.
\newblock {\em Journal of the Royal Statistical Society: Series B (Statistical
  Methodology)}, 71(2):319--392.

\bibitem[Rue et~al., 2011]{inlapkg}
Rue, H., Martino, S., and Chopin, N. (2011).
\newblock {\em INLA: Approximate Bayesian Inference using Integrated Nested
  Laplace Approximations}.

\bibitem[Schmidt and Gelfand, 2003]{schmidt2003}
Schmidt, A.~M. and Gelfand, A.~E. (2003).
\newblock A bayesian coregionalization approach for multivariate pollutant
  data.
\newblock {\em Journal of Geophysical Research: Atmospheres}, 108(D24).

\bibitem[Sharabiani and Mahani, 2017]{cfcpkg}
Sharabiani, M.~T. and Mahani, A.~S. (2017).
\newblock {\em CFC: Cause-Specific Framework for Competing-Risk Analysis}.

\bibitem[Simpson et~al., 2017]{simpson2017}
Simpson, D., Rue, H., Riebler, A., Martins, T.~G., S{\o}rbye, S.~H., et~al.
  (2017).
\newblock Penalising model component complexity: {A} principled, practical
  approach to constructing priors.
\newblock {\em Statistical Science}, 32(1):1--28.

\bibitem[Van~Niekerk et~al., 2019]{van2019}
Van~Niekerk, J., Bakka, H., and Rue, H. (2019).
\newblock Joint models as latent gaussian models-not reinventing the wheel.
\newblock {\em arXiv preprint arXiv:1901.09365}.

\bibitem[Williamson et~al., 2008]{williamson2008}
Williamson, P., Kolamunnage-Dona, R., Philipson, P., and Marson, A. (2008).
\newblock Joint modelling of longitudinal and competing risks data.
\newblock {\em Statistics in medicine}, 27(30):6426--6438.

\end{thebibliography}

\end{document}